\pdfoutput=1
\documentclass[conference]{IEEEtran}
\IEEEoverridecommandlockouts
\newlength{\survcol}

\usepackage{graphicx,url}
\usepackage[utf8]{inputenc}
\usepackage[english]{babel}
\usepackage[T1]{fontenc}
\usepackage{amssymb}
\usepackage{amsmath}
\usepackage{booktabs}
\usepackage{array}
\usepackage{tabularx}
\usepackage{enumitem}
\usepackage{pifont}
\usepackage[table]{xcolor}
\usepackage{soul} 
\setulcolor{green}  
\let\soulul\ul    
\renewcommand{\ul}[1]{\textcolor{blue}{\soulul{#1}}} 
\usepackage{calc}
\usepackage{float}
\usepackage[hidelinks]{hyperref}
\usepackage{listings}
\usepackage{rotating}
\usepackage[numbers,sort&compress]{natbib}
\usepackage{wasysym}
\usepackage{stfloats}
\usepackage{longtable}
\selectfont
\definecolor{rowalt}{gray}{0.96}   
\definecolor{qband}{gray}{0.90}    

\newcommand{\tdef}[3]{\textbf{#1} & \textsc{#2} & #3 \\}

\newcommand{\NA}{---}

\definecolor{cmdcomment}{gray}{0.45}
\lstdefinestyle{cmds}{
  basicstyle=\scriptsize\ttfamily,
  commentstyle=\color{cmdcomment}\itshape,
  language=bash,
  morecomment=[l]{\#},
  frame=none,
  breaklines=true,
  breakindent=0pt,
  columns=fullflexible,
  keepspaces=true,
  xleftmargin=1em,
  aboveskip=5pt,
  belowskip=5pt,
  showstringspaces=false,
}
\newcommand{\ubheaderA}{\textbf{Sandbox} & \textbf{T1} & \textbf{T2} & \textbf{T3} & \textbf{T4} & \textbf{T5} & \textbf{T6} & \textbf{T7} & \textbf{T8} & \textbf{T9} & \textbf{Pass} \\}
\newcommand{\ubheaderB}{\textbf{Sandbox} & \textbf{T1} & \textbf{T2} & \textbf{T3} & \textbf{T4} & \textbf{T5}  & \textbf{Pass} \\}
\newcommand{\ubheaderC}{\textbf{Sandbox} & \textbf{T1} & \textbf{T2} & \textbf{T3} & \textbf{T4} & \textbf{T5} & \textbf{T6} & \textbf{T7} & \textbf{T8} & \textbf{T9}  & \textbf{Pass} \\}
\newcommand{\ubbodyA}{%
Docker Sbx & \yes &\yes & \no & \yes & \no & \no  & \yes & \yes & \yes &  6 \\
Microsandbox & \yes & \no & \no & \yes & \no & \no & \yes & \yes & \yes & 5 \\
Shuru & \yes & \yes & \no & \no & \no & \no & \yes & \yes & \yes & 5 \\
SmolVM &  \yes & \no & \no & \yes & \no & \no & \yes & \yes & \yes & 5 \\
VibeBox & \NA & \NA & \NA & \NA & \NA & \NA & \NA & \NA & \NA & \NA \\
\midrule
\textbf{SideKernel} & \yes & \yes & \no & \yes & \yes & \yes & \yes & \yes & \yes & 8 \\
}

\newcommand{\ubbodyB}{%
Docker Sbx &  \yes  & \yes & \yes & \yes & \yes  & 5 \\
Microsandbox  & \no & \yes & \no & \yes & \yes & 3\\
Shuru &  \no & \yes & \no & \yes & \yes & 3 \\
SmolVM  & \no & \yes & \no & \yes & \yes  & 3 \\
VibeBox & \NA & \NA & \NA & \NA & \NA  & \NA \\
\midrule
\textbf{SideKernel} & \yes & \yes & \no & \yes & \no &  3 \\
}

\newcommand{\ubbodyC}{%
Docker Sbx & \yes & \no & \yes & \yes & \yes & \yes & \yes &  \yes & \yes & 8 \\
Microsandbox & \no & \no & \yes  & \yes & \yes & \yes & \no  & \no & \no & 4 \\
Shuru & \no & \no & \yes & \yes & \yes & \yes & \no & \no & \no &  4 \\
SmolVM & \no & \no & \yes & \yes & \yes & \yes & \no & \no  & \no  & 4 \\
VibeBox & \NA & \NA & \NA & \NA & \NA & \NA & \NA & \NA & \NA  & \NA \\
\midrule
\textbf{SideKernel} & \yes & \yes & \yes & \yes & \yes & \yes & \yes & \yes & \no &   8 \\
}

\newcommand{\ubtable}[7]{%
\begin{table}[H]
\centering
\caption{#2}
\label{#1}
\setlength{\tabcolsep}{1.5pt}
\renewcommand{\arraystretch}{1.05}
\scriptsize
\rowcolors{2}{rowalt}{}
\begin{tabularx}{\columnwidth}{>{\raggedright\arraybackslash}p{#6} *{#3}{>{\centering\arraybackslash}X} !{\vrule width 0.6pt} >{\centering\arraybackslash}p{0.55cm}}
\toprule
#4
\midrule
#5
\bottomrule
\end{tabularx}
\vspace{2pt}
\vspace{2pt}
\rowcolors{1}{rowalt}{}
\begin{tabularx}{\columnwidth}{@{}l@{\hspace{4pt}} >{\raggedright\arraybackslash}X@{}}
#7
\end{tabularx}
\end{table}}

\newcommand{\yes}{\ding{51}}
\newcommand{\no}{\ding{55}}

\definecolor{valGreen}{HTML}{2E7D32}
\definecolor{varRed}{HTML}{C62828}
\definecolor{varBlue}{HTML}{1555C0}
\newcolumntype{L}[1]{>{\raggedright\arraybackslash}p{#1}}
\newcolumntype{C}[1]{>{\centering\arraybackslash}p{#1}}
\newcolumntype{R}[1]{>{\raggedleft\arraybackslash}p{#1}}

\newcolumntype{N}{>{\raggedright\arraybackslash\hsize=0.83\hsize}X}  
\newcolumntype{S}{>{\centering\arraybackslash\hsize=0.76\hsize}X}     
\newcolumntype{M}{>{\raggedright\arraybackslash\hsize=1.97\hsize}X}   
\newcolumntype{K}{>{\raggedright\arraybackslash\hsize=1.12\hsize}X}   
\newcolumntype{A}{>{\raggedright\arraybackslash\hsize=2.18\hsize}X}
\newcolumntype{B}{>{\centering\arraybackslash\hsize=0.64\hsize}X}
\newcolumntype{G}{>{\centering\arraybackslash\hsize=0.90\hsize}X}

\newcolumntype{P}{>{\raggedright\arraybackslash\hsize=1.92\hsize}X}
\newcolumntype{Q}{>{\centering\arraybackslash\hsize=0.64\hsize}X}
\newcolumntype{Z}{>{\centering\arraybackslash\hsize=0.80\hsize}X}

\newcolumntype{U}{>{\raggedright\arraybackslash\hsize=2.48\hsize}X}
\newcolumntype{V}{>{\centering\arraybackslash\hsize=0.62\hsize}X}
\newcolumntype{Y}{>{\centering\arraybackslash\hsize=1.03\hsize}X}

\title{SideKernel: A Usable microVM Sandbox for AI Coding Agents on macOS}

\author{
\IEEEauthorblockN{Dimitrios Prasakis}
\IEEEauthorblockA{Georgia Institute of Technology\\
dimitrios.prasakis@gatech.edu}}

\begin{document}

\maketitle

\begin{abstract}
AI coding agents are untrusted system components, yet they require autonomy on the developer machines they run on. This contradiction is a security problem. Sandboxes provide an isolated environment, but for local macOS development, the existing local, open-source options for AI coding agents are few in number and cumbersome to use. I conducted a formative online user survey which indicates that fewer than 40\% of AI coding agent users run their agents in a sandbox and identifies the top usability barriers hindering AI coding agent sandbox adoption. These findings are used to develop SideKernel: an open-source, local, microVM-based macOS sandbox for AI coding agents designed for usability. To evaluate SideKernel, I compiled a list of sandboxes available on the market and filtered it against five inclusion criteria. Then I performed a comparative analysis between SideKernel and the sandboxes that satisfy these criteria, across 23 capability tests derived from the usability barriers revealed by the user survey. I discovered that only a few sandboxes are similar to SideKernel, and that among those, Docker Sandboxes and SideKernel score highest on capability features related to usability. A secondary contribution of this paper is a survey of the existing solution space for local, open-source, microVM-based macOS sandboxes for AI coding agents.
\end{abstract}


\section{Introduction}
\label{sec:introduction}

\subsection{Motivation}

AI agents are untrusted system components, and running untrusted system components in a sandbox is a recommended practice in systems security~\cite{christodorescu2026agent}. Over the past year\footnote{Approximated from the date of the first public commit of most sandboxes listed in Table~\ref{tab:inclusion}.}, the cybersecurity market has seen rapid development of AI agent sandboxes~\citep{nono, landrun, fence, safehouse, cellmate, daytona, yolobox, aio, denosandbox, capsule, pctx, e2b, microsandbox, shuru, vibebox, cua, smolvm, sprites, arrakis, opensandbox, docker-sbx, coderunner}---many of which have gained popularity but, anecdotally, see limited adoption with the majority of AI coding agent users not running a sandbox (Section~\ref{sec:survey}).  The implication of this finding is that most users run AI coding agents on their machines, with all the security risks that this implies. For example, AI agents are vulnerable to prompt injection attacks, tool misuse leading to remote code execution, data exfiltration, and the installation of malware through hallucinated package names \cite{owasp2025agentic, spracklen2025package}.

\subsection{Isolation Primitives}

The isolation primitives of sandboxes can be broadly classified based on where the security boundary lives \citep{agache2020firecracker}:

\begin{enumerate}[leftmargin=1.4em,itemsep=3pt,topsep=4pt]
    \item \textbf{OS-isolation (Shared Kernel):} Isolation primitives that share the host's kernel. For example, Linux Containers, bubblewrap (namespaces \& cgroups), or kernel-enforced security controls such as Linux Landlock, macOS Seatbelt/sandbox-exec, and others.
    \item \textbf{Language-Specific Isolation:} Isolation primitives where the boundary lives inside a managed runtime provided by the language. Examples include WASM runtimes, Chromium's site isolation, and the JVM.
    \item \textbf{Virtualization (Separate Kernel):} Each workload runs on its own dedicated kernel, whether a full virtual machine such as QEMU/KVM, VirtualBox, or Parallels, a unikernel \cite{unikraft} or a microVM (i.e., a stripped-down guest running a minimal Linux kernel) such as Firecracker or Kata Containers.
\end{enumerate}

\subsection{Scope}

Many of the AI coding agent sandboxes on the market use isolation primitives that rely on a shared kernel. Container-based isolation is weaker than microVM isolation \cite{agache2020firecracker}. This claim is, however, not trivial to establish for other primitives such as macOS Seatbelt, which on the one hand is a shared kernel isolation primitive, but on the other hand is policy-based and enforced at the kernel level. The right question to ask is: under which threat model do I evaluate the security of the two? This is a research question that deserves a study of its own, and is out of scope for this paper. However, I state in Section~\ref{sec:threatmodel} the threat model under which SideKernel is designed.

Moreover, security depends not only on the theoretical guarantees of a technology, but also on the actual implementation of it. In this paper, I focus exclusively on microVM sandboxes. The security argument is architectural: the host's kernel is not reachable, as the untrusted agent interacts with a stripped-down microVM Linux kernel whose attack surface is by definition reduced. A malicious agent or adversary would have to escape the virtual machine monitor (VMM) to land on the host. This is the isolation primitive that many cloud providers use for untrusted code. For example, AWS uses Firecracker microVMs for Lambda and Fargate~\cite{agache2020firecracker}, as do the major commercial hosted AI agent sandboxing platforms such as E2B and Vercel Sandbox~\cite{e2b,vercel-sandbox}.

\subsection{Research Question and Contributions}

This paper addresses the following research question:

\begin{quote}
\textit{Which usability barriers hinder macOS developers' adoption of AI coding agent sandboxes?}
\end{quote}

In order to answer this question, this paper makes three contributions:

\begin{enumerate}[leftmargin=1.4em,itemsep=3pt,topsep=4pt]
  \item \textbf{SideKernel:}  An implementation of a local, microVM-based macOS sandbox for AI coding agents. SideKernel's core thesis is usable security: a sandbox that is secure, easy to use, and built exclusively for AI coding agents.
  \item \textbf{An empirical account of sandbox usability issues:} Informal formative research via an online user Google survey that hints at the \textsc{usability barriers} that AI coding agent users face when using similar sandboxes. This is a secondary contribution of this paper that ties to the development of SideKernel; however, this study should not be treated as statistically generalizable, but rather as an informal signal. More on this in Section~\ref{sec:limitations}.
  \item \textbf{A comparative analysis of the existing sandbox market:} Summative research via a comparative analysis of SideKernel against a list of similar sandboxing solutions on the market. This is a secondary contribution of this paper that discovers sandboxes similar to SideKernel and compares them.
\end{enumerate}

\section{Background}
\label{sec:background}

\subsection{Terminal-Based vs.\ Programmatic AI Agents}

AI coding agents are typically executed in the terminal and require continuous human interaction. Examples include Claude Code \cite{anthropic2026claudecode}, Codex \cite{openai2026codex}, and others. Programmable software development kit (SDK) agents are typically executed on a server, and require little to no human interaction. They are programs interacting with MCP servers \cite{mcp-definition} or directly with the application programming interfaces of LLM providers or local models through their official SDKs. This distinction matters because this paper is concerned only with AI coding agents and their related sandboxes.

\subsection{Threat Model}
\label{sec:threatmodel}

AI coding agents are untrusted system components and they are executed directly on the host's terminal. Therefore, SideKernel's threat model treats the agent itself as the adversary. For example, an agent might be prompt-injected from external sources, directly or indirectly, or might hallucinate dangerously during its own inference. The
reason the agent turns hostile is immaterial to the threat model: the model treats the threat actor as fully privileged (inside the sandbox) regardless of cause.

\begin{figure}[!t]
    \centering
    \includegraphics[width=1\columnwidth]{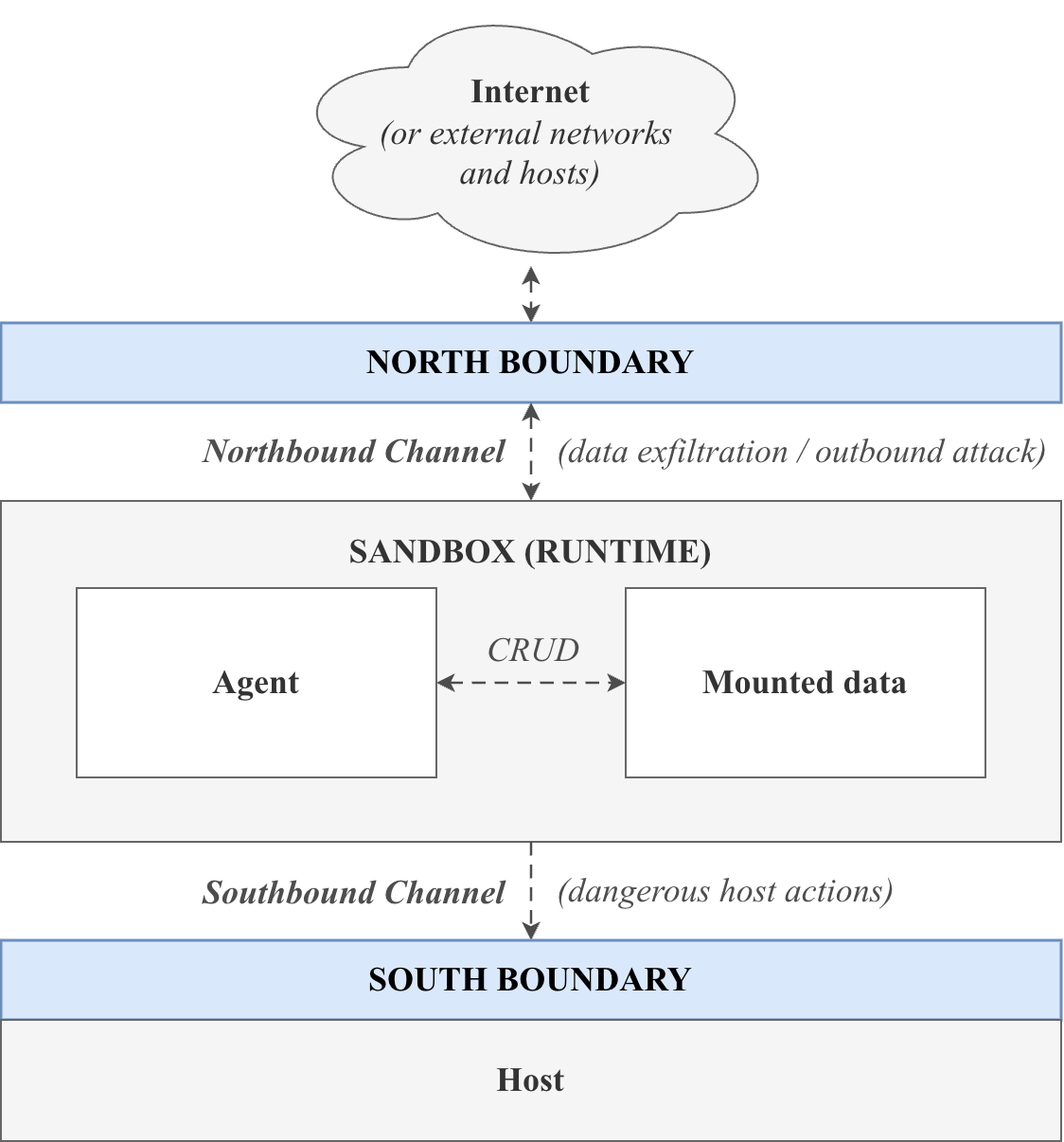}
    \caption{North \& South Security Boundaries}
    \label{fig:boundaries}
\end{figure}

To better study the attack surface of AI agents in the context of sandboxing, I define two security boundaries, depicted in Fig.~\ref{fig:boundaries}:

\begin{enumerate}
    \item \textsc{North Boundary:} The north boundary encompasses all communications that the agent inside the sandbox conducts with machines other than the host or localhost. This includes the internet, or other hosts in the same or external networks. I refer to the communication channel from the sandbox to the north boundary as \textsc{Northbound Channel}.
    \item \textsc{South Boundary:} The south boundary includes all communications that the sandbox performs with the host. Depending on the exact implementation of the sandbox, that could be over virtual sockets, or networking protocols such as SSH. I refer to the communication channel from the sandbox to the host boundary as \textsc{Southbound Channel}.
\end{enumerate}

In this context, the south boundary exists to \textbf{protect the host} from malicious actions, for example installing malware, or deleting or editing valuable files. The threat is unidirectional: from the sandbox towards the host.
The north boundary, in turn, \textbf{protects the data} that is mounted inside the sandbox from exfiltration, for example a malicious agent sending all the mounted data to an external source. It is equally concerned with \textbf{protecting external entities from a local malicious agent}. In this context, the threat is bidirectional (from the sandbox to the north boundary and vice versa).

A notable example is OpenAI's autonomous AI agent system inadvertently hacking Hugging Face through a sandbox escape \cite{openai2026huggingface}. That very example, however, also shows the complexity of the enterprise of sandboxing powerful agents. OpenAI's agent bypassed northbound security controls by discovering and exploiting a chain of vulnerabilities to eventually escape the sandbox and local networks and reach the internet.

The security controls of SideKernel address both boundaries, though not equally. 

\subsubsection{South Boundary Threat Model}
On the south boundary, the goal is protecting the \textit{host's runtime} and its \textit{non-mounted filesystem} from adversarial agents running inside the sandbox.
\subsubsection{North Boundary Threat Model}
On the north boundary, the goal is preventing \textit{exfiltration of sandbox data} to the internet (or other external networks/endpoints), as well as using the sandbox as a \textit{launch point} for outbound offensive operations by a malicious agent.

\section{Related Work}
\label{sec:related}

\subsection{Existing Sandboxes}

The current sandboxing market is dominated by sandboxes that rely on OS-isolation or
language-specific isolation. This includes the native sandboxes of major
frontier-model companies, for example, Claude Code~\cite{claude-sandbox},
Codex~\cite{codex-sandbox}, Gemini~\cite{gemini-sandbox}, and Grok
Build~\cite{grok-sandbox}---as well as well-known open-source tools that depend on Seatbelt and Landlock, such as Nono~\cite{nono}, Agent
Safehouse~\cite{safehouse}, and
more~\citep{yolobox, capsule, pctx, aio}.

Others use hardware-virtualized microVMs, but they are SaaS products with no
local deployment, indicatively E2B~\cite{e2b} and Daytona~\cite{daytona}; are not
open-source~\citep{docker-sbx}; are general VM runtimes rather than purpose-built
agent sandboxes~\citep{lima, apple-container, orb}; focus primarily on computer
use~\cite{cua}; or target mainly programmable SDK agents or MCP gateways rather than
terminal-based coding agents~\citep{microsandbox, arrakis}. 

To the best of my knowledge, SmolVM~\cite{smolvm}, Microsandbox~\cite{microsandbox}, Shuru~\cite{shuru}, and VibeBox~\cite{vibebox} are the only open-source, local, purpose-built for terminal-based AI coding agents,
hardware-virtualized with microVMs, and macOS-native sandboxes on the market. I refer to these five
properties as the \textsc{inclusion criteria} (IR-1--IR-5) and define and discuss their importance in Section~\ref{sec:comparative_analysis_methodology}. 

This is not an exhaustive survey of all the sandboxes on the market. Rather, it aims to present an overview of the different types of sandboxes that exist, but more importantly identify a complete list of all the sandboxes similar\footnote{What constitutes a similar sandbox is discussed in detail in Section~\ref{sec:comparative_analysis_methodology}.} to SideKernel.

\section{Methodology}
\label{sec:methodology}

SideKernel's prototype implementation is informed by two studies:

\begin{enumerate}
    \item A \textbf{user survey (formative research)} \citep[Ch.~3.1.1]{albert2022measuring} conducted prior to the development of SideKernel (complete instrument in Table~\ref{tab:survey}). The purpose of this formative online user survey is twofold: 
    \begin{enumerate}
        \item to inform the implementation of SideKernel
        \item to motivate a comparative analysis of SideKernel against other sandboxes on the market.
    \end{enumerate} 
\item A \textbf{comparative analysis (summative research)} \citep[Ch.~3.1.2]{albert2022measuring} benchmarking the final SideKernel prototype against other market sandboxes that satisfy a set of inclusion criteria (Table~\ref{tab:inclusion} and Fig.~\ref{fig:IR_filtering}). The analysis is run on a set of well-defined capability tests (Section~\ref{sec:evaluation}), grounded on the usability signals of the user survey.
\end{enumerate}

\begin{figure}[!t]
    \centering
    \includegraphics[width=1\columnwidth]{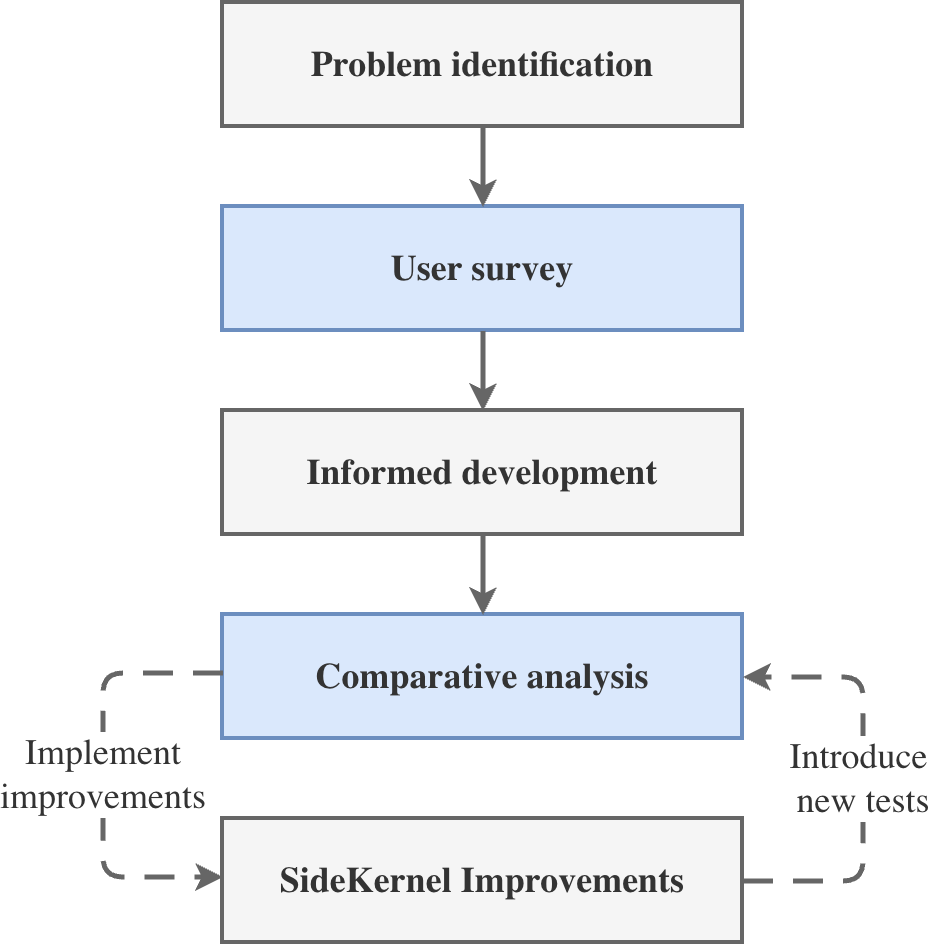}
    \caption{SideKernel research methodology.}
    \label{fig:methodology}
\end{figure}

The depiction of the methodology for developing SideKernel is shown in Fig.~\ref{fig:methodology}, where a feedback loop between the results of the comparative-analysis and the continuous development of SideKernel ensures that the software is constantly improved, by addressing the findings and adding new tests. The goal of this methodology is to allow me to evaluate and improve SideKernel continuously.

\subsection{User Survey Methodology}
\label{sec:user_survey_methodology}

The user survey takes the form of an online Google Survey that uses a mix of multiple-choice, multi-select, and free-text questions, gathering empirical evidence from sandbox users on what \textsc{usability barriers} they faced while using similar sandboxes.

The user survey is structured in three steps. First, filtering out users that have never used an AI coding agent. Second, categorizing the remaining users based on whether they are current or former sandbox users. Third, through multiple-choice, multi-select, and free-text questions, I gathered evidence on what usability barriers they faced while using those sandboxes.

Respondents were gathered via direct and targeted recruiting \citep[Ch.~2]{albert2010beyond} specifically through my direct professional network, Zurich's JFloor startup community\footnote{\url{https://thejfloor.com/}} and Georgia Tech's practicum course (CS-6727) online forum and user group \& Georgia Tech's Slack channels. It is important to note that this is an informal Google Survey. I treat the results as anecdotal signals rather than statistically generalizable results.  

\subsection{Comparative Analysis Methodology}
\label{sec:comparative_analysis_methodology}

 The goal of the comparative analysis is to examine how well the final SideKernel prototype addresses the usability barriers identified from the user study relative to other sandboxes currently on the market. 

The comparative analysis begins with the compilation of a list of all existing sandboxes that are similar to SideKernel. To define what similarity means, I defined a set of \textsc{inclusion criteria} that I filter the list against. These are the following:

\begin{enumerate}
    \item \textsc{Open-Source}: The sandbox must be free to use, and either open-source or source-available, to foster trust through a publicly inspectable codebase. 
    \item \textsc{Local}: The sandbox must execute in the user's own machine, rather than in a remote execution environment such as a cloud or a virtual private server (VPS).
    \item \textsc{Interactive shell}: The sandbox must provide an interactive shell (e.g., bash) intended for operating CLI coding agents such as Claude, Codex etc. This excludes sandboxes that \textit{only} provide SDK interfaces or computer-use interfaces (GUI) for AI agents.
    \item \textsc{Hardware-virtualization (microVM)}: The sandbox must use hardware virtualization as its isolation boundary, and provide this virtualization performantly through a stripped down kernel (microVM or unikernels). This excludes platforms intended for full VM virtualization, as performance is not a named but an implied requirement.
    \item \textsc{macOS Support}: The sandbox must be able to run on a macOS ARM machine.
\end{enumerate}

\begin{table*}[!tb]
\centering
\caption{Inclusion requirements and isolation classification. IR-1: open-source; IR-2: local execution; IR-3: interactive shell for CLI agents; IR-4: hardware-virtualized; IR-5: macOS}
\label{tab:inclusion}
\setlength{\tabcolsep}{2pt}
\renewcommand{\arraystretch}{1.3}
\scriptsize
\rowcolors{3}{rowalt}{}
\begin{tabularx}{\textwidth}{N *{5}{S} !{\vrule width 0.8pt} M K}
\toprule
& \multicolumn{5}{c}{\textbf{Inclusion requirements}}
 & \multicolumn{2}{c}{\textbf{Isolation}} \\
\cmidrule(lr){2-6}\cmidrule(l){7-8}
\textbf{Sandbox}
 & \rotatebox{55}{IR-1 Source}
 & \rotatebox{55}{IR-2 Local}
 & \rotatebox{55}{IR-3 Shell}
 & \rotatebox{55}{IR-4 HW-virt.}
 & \rotatebox{55}{IR-5 macOS}
 & \textbf{Isolation Mechanism(s)} & \textbf{Class} \\
\midrule
nono             & \yes & \yes & \yes & \no  & \yes & Seatbelt / Landlock         & Shared Kernel \\
Landrun          & \yes & \yes & \yes & \no  & \no  & Landlock                    & Shared Kernel \\
Fence            & \yes & \yes & \yes & \no  & \yes & Seatbelt / bubblewrap       & Shared Kernel \\
Daytona          & \yes$^{a}$ & \no  & n/a$^{b}$ & n/a$^{b}$ & n/a$^{b}$ & Shared Kernel / microVM & Cont.\ / Sep.\ kernel \\
Yolobox          & \yes & \yes & \yes & \no  & \no$^{c}$ & Docker/Podman container & Shared Kernel \\
AIO              & \yes & \yes & \no  & \no  & \no$^{c}$ & Docker container        & Shared Kernel \\
Deno             & \no  & \no  & n/a$^{b}$ & n/a$^{b}$ & n/a$^{b}$ & microVM  & Sep.\ kernel \\
Capsule          & \yes & \yes & \no  & \no  & \yes & WASM              & Runtime \\
pctx             & \yes & \yes & \no  & \no  & \yes & Deno/V8 isolate; Podman     & Runtime / Cont. \\
E2B              & \yes & \no  & n/a$^{b}$ & n/a$^{b}$ & n/a$^{b}$ & microVM  & Sep.\ kernel \\
Microsandbox     & \yes & \yes & \yes & \yes & \yes & microVM       & Sep.\ kernel \\
Shuru            & \yes & \yes & \yes & \yes & \yes & microVM         & Sep.\ kernel \\
VibeBox          & \yes & \yes & \yes & \yes & \yes & microVM        & Sep.\ kernel \\
CUA              & \yes & \yes & \no$^{d}$ & \yes & \yes & VM   & Sep.\ kernel \\
SmolVM           & \yes & \yes & \yes & \yes & \yes & microVM            & Sep.\ kernel \\
Sprites          & \no  & \no  & n/a$^{b}$ & n/a$^{b}$ & n/a$^{b}$ & microVM  & Sep.\ kernel \\
Arrakis          & \yes & \yes & \yes & \yes & \no  & microVM  & Sep.\ kernel \\
OpenSandbox      & \yes & \yes & \no  & \yes$^{e}$ & \no$^{c}$ & Container (opt. MicroVM)  & Shared Kernel \\
Docker Sbx       & \no  & \yes & \yes & \yes & \yes & microVM   & Sep.\ kernel \\
CodeRunner       & \yes & \yes & \no$^{f}$ & \yes & \yes & microVM   & Sep.\ kernel \\
Agent Safehouse       & \yes & \yes & \yes & \no & \yes & Seatbelt   & Shared kernel \\
\midrule
\textbf{SideKernel} & \yes & \yes & \yes & \yes & \yes & microVM     & Sep.\ kernel \\
\bottomrule
\end{tabularx}
\smallskip
\raggedright\footnotesize
\newline
$^{a}$ Since June 2026, Daytona moved into private repositories. \\
$^{b}$ Not applicable: the sandbox executes on remote infrastructure rather than on the user's machine (fails IR-2), so the remaining requirements, which concern the local macOS execution environment, are not evaluated. \\
$^{c}$ Installable on macOS, but the sandbox does not virtualize natively: it executes only inside a Linux VM supplied by Docker Desktop, Colima, or Podman machine. \\
$^{d}$ CUA is built for computer use. It is heavy (boots full macOS VMs) and has a different purpose (GUI, computer use) than SideKernel. \\
$^{e}$ Evaluated in its default configuration (shared host kernel). \\
$^{f}$ No native shell subcommand; interactive access uses the apple/container substrate.
\end{table*}

Sandboxes that meet inclusion criteria two to five are included in the comparative study. I did not filter on open-source, because I did not want to exclude the major player, Docker Sandboxes.

\begin{figure}[!t]
    \centering
    \includegraphics[width=1\columnwidth]{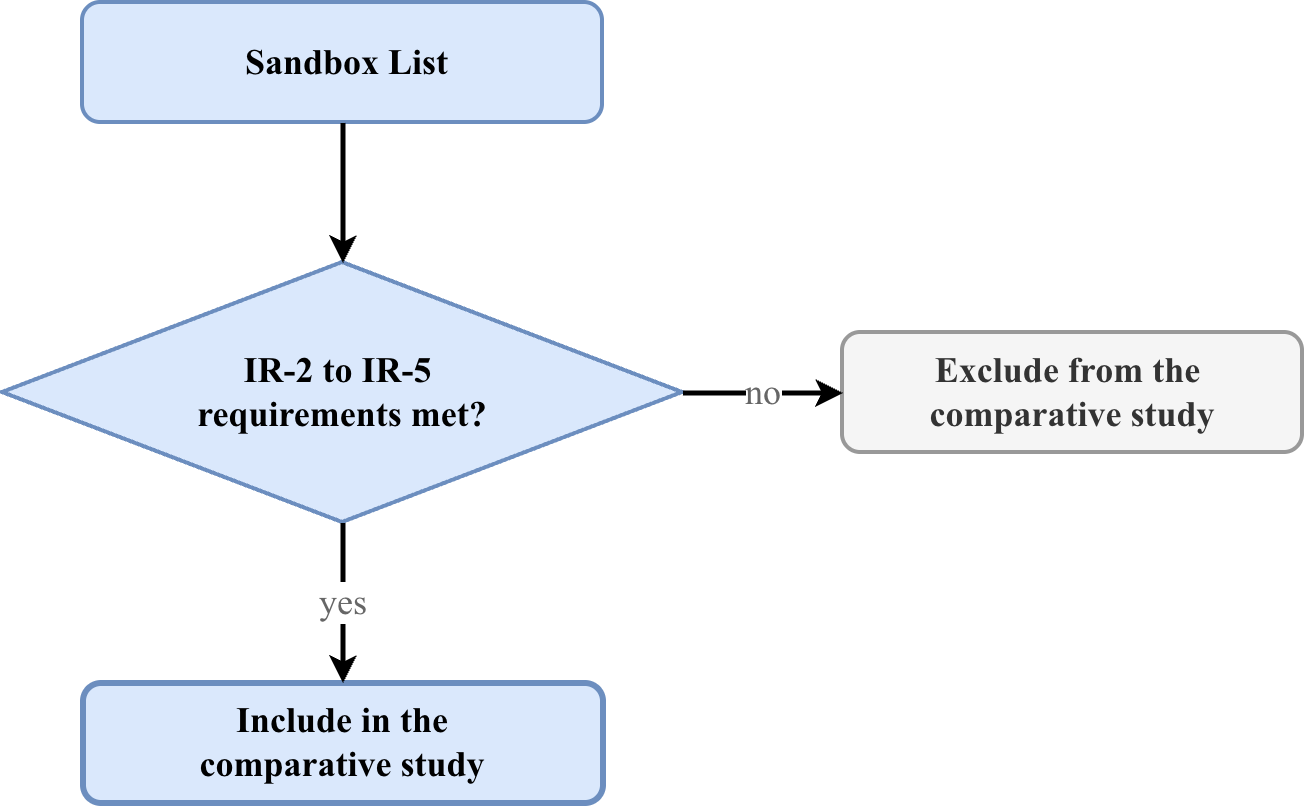}
    \caption{Filtering process against inclusion criteria.}
    \label{fig:IR_filtering}
\end{figure}

\subsection{Deriving Tests from the Usability Barriers}
\label{sec:deriving_barriers}

The user survey, however, gives little information about the detailed tests for each barrier. For example, the user study suggests that sharing files between the host and the sandbox is something users find cumbersome. But what exactly does this mean? To the best of my knowledge, there is no prior literature that studies these attributes in any greater detail for sandboxes, beyond this study.

The selection of the tests for each usability barrier is done so that as many possibilities as possible are covered---while making reasonable assumptions. For example, in file sharing (UB-1), I noticed that mounting is common across sandboxes, but file sharing over \texttt{SFTP} or \texttt{SSH} is not; hence I created technical tests around mounting. Indicatively, assessing whether CRUD operations work bidirectionally, if file-bit flags (r, w, x) transfer, and more. Another example is image clipboard sharing, a feature supported by coding agents such as Claude.

The multi-select and multiple-choice answers are then grouped by frequency to identify the top usability barriers. For example the answers ``Initial installation / setup is too hard'' and ``Had to set everything up from scratch (dotfiles, shell, tools, logins)'' are merged into one category ``Environment provisioning \& Setup''.

The comparative analysis provides an overview of specific usability capabilities within the confines of the tested usability barriers, rather than a generalizable statement about which sandbox is more useful.

\subsection{Benchmark Construction}
\label{sec:benchmark_construction}

Each task is a \textsc{capability test} with a binary outcome, testing whether an action related to a specific usability barrier is feasible in a given sandbox. Success criteria for every capability test are fixed before the comparative analysis iteration (evaluation) begins~\citep[\S4.1]{albert2022measuring}.

A test passes if and only if the sandbox provides a \textsc{built-in mechanism} for the task. By built-in mechanism I refer to a documented feature of the sandbox itself: one discoverable through the tool's official documentation or its own interface. A test fails if the task is infeasible, or feasible only by working around the sandbox---for example blocking network access to a sandbox by explicitly and manually blocking the sandbox's networking on a host's firewall. More specifically, the binary outcomes of the capability tests are defined as follows: 
\begin{itemize}[leftmargin=1.4em,itemsep=3pt,topsep=4pt]
	\item \yes~ = a built-in mechanism for the task exists
	\item \no~ = no built-in mechanism exists
\end{itemize}

This evaluation rule is deliberately generous, and deliberately blind to how easy or hard the built-in mechanism is to use. Because I am both the tester and developer of SideKernel, I had to find ways to reduce bias in my evaluation as much as possible. One way to do this is to have a standardized methodology to test the sandboxes. Had I also scored on perceived effort, then every cell would encode a subjective judgment of my own, and the comparison would not be reproducible by a reader. Restricting the binary outcome to built-in capabilities makes the scores objective and auditable.

A valid counterargument to this methodology is that this capability binary scoring measures exactly that: capability, not usability. While this statement is true, there is a nuance: these capabilities are sourced from the user study which identifies which usability barriers hinder sandbox adoption. \textit{Therefore they are capabilities related to usability barriers.}

But this alone is not sufficient to demonstrate usability on its own. Just because a sandbox has built-in capabilities to perform some actions---even when those actions are related to solving usability barriers---it does not mean that this built-in mechanism is intuitive and frictionless. To counter this transparently, I therefore record \textbf{friction notes} alongside the binary outcomes. Friction notes never change a binary outcome. They are reported per barrier in Section~\ref{sec:evaluation} and carry the qualitative descriptions that the binary outcome cannot, thus allowing the reader to make a judgment of their own.

The sum of a sandbox's binary outcomes forms its \textbf{capability score}. It is a naive, unweighted sum in which every test counts equally, although in practice not every task is equally consequential. It is a signal rather than a definitive measure, and it should be read together with the friction notes rather than on its own.

\subsection{Evaluation Protocol}
\label{sec:evaluation_protocol}

In order to make the comparison standard and as thorough as possible, I create a template directory that contains the following structure and files:

\begin{lstlisting}[language=, basicstyle=\footnotesize\ttfamily, frame=none, breaklines=true, breakindent=0pt]
sandbox-test-template/
|-- external-file.txt           # host only file
|-- big-file.txt                # 10 GB file  
`-- mount/                      # mounted dir
    |-- requirements.txt        # contains numpy 
    |-- workspace-file.txt      # CRUD, symlink 
    `-- shell.sh                # no exec-bit 
\end{lstlisting}

This directory is deployed fresh for each new sandbox evaluation. 

These tests are extensive, but not exhaustive---and should not be read as such. They are, however, a reasonable and transparent indicator of how capable and usable different sandboxes are in relation to the barriers the user survey uncovered.

Another important aspect of this evaluation methodology is that the iterative approach of SideKernel's development (Fig.~\ref{fig:methodology}) allows me to continuously create new tests or amend existing ones while improving the product at the same time.

\section{User Survey Results}
\label{sec:survey}

\subsection{Google Survey Report Analysis}

Below I present the most frequently mentioned user survey results that shaped my decisions on the development of the prototype of SideKernel, as well as its evaluation. The complete instrument and responses are reported in
Appendix~\ref{app:survey}. The free-text responses are only analyzed and reported in aggregate in this section.

\begin{itemize}[leftmargin=1.4em,itemsep=4pt,topsep=4pt]
    \item The Google survey gathered a total of 80 respondents.
    \item Out of the 80 respondents, 68 use AI coding agents (85\%). Respondents who do not use AI coding agents are filtered out and not considered for the following results.
    \item 60.3\% of AI coding agent users are macOS users.
    \item Out of the 68 AI coding agent users, 86.8\% use Claude Code (59 respondents), 38.2\% use Codex (26 respondents), and 33.8\% use Microsoft's Copilot in agent mode (23 respondents). Cursor (19.1\%, 13 respondents) and Gemini CLI (14.7\%, 10 respondents) follow, while every remaining agent sits at or below 2.9\%, making them individually insignificant for the scope of SideKernel.
    \item Only 38.2\% (26 users) of respondents are current sandbox users, while 10.3\% (7 users) are former sandbox users and 51.5\% (35 users) have never used a sandbox.
    \item From the current sandbox users (26 in total), 34.6\% (9 users) wrote that they face no significant difficulties.
    \item From the former and current sandbox users (33 in total), 8 wrote that getting ports / localhost / networking to work is difficult, and that having to set everything up from scratch (e.g., dotfiles, shell, tools, logins) is also difficult; 7 wrote that sharing files between the sandbox and the host is hard or slow. They are followed by 5 users noting that multi-sandbox management is confusing and that initial setup and installation was confusing.
    \item Notably, among the non-sandbox users, 60\% answered that they have never considered it and 25.7\% that the security benefit of using a sandbox did not seem worth the effort for their use case. 
\end{itemize}

The sample is highly technical, with one third working in security and the vast majority being engineers. The dominance of Claude Code helped me scope down my practicum and focus only on Claude Code. The count of 33 former and current users is the group of interest for this study.

The free-text responses of current and former users around what features or improvements an ideal sandboxing solution would offer compared to their current or past setups focus on a few main points. In aggregate users ask for ease of use and ``nearly invisible'' sandboxes that require minimum effort to manage and operate. Another cluster of users emphasizes the importance of reproducible provisioning. That includes features such as saving a working setup as a template and flexibility to add tools.  A notable response described a ``golden image'' preloaded with tools; that gave me the idea for the personalized image  SideKernel implements today. Finally, many comments note that
granularity in the policies for the security controls, as well as observability, is an important attribute.

\begin{figure}[!t]
    \centering
    \includegraphics[width=\columnwidth]{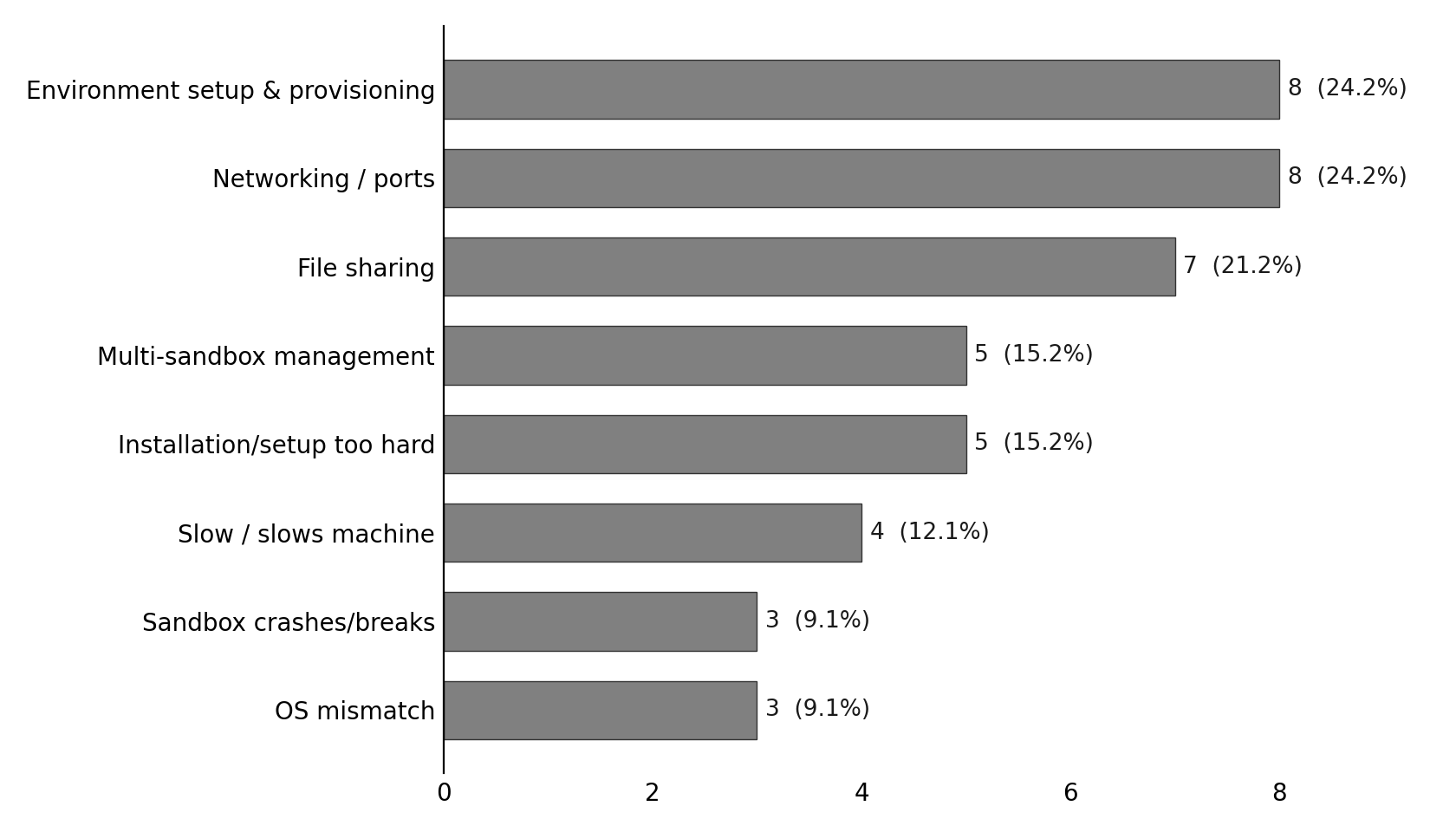}
    \caption{Usability barriers reported by current and former sandbox users ($n=33$).}
    \label{fig:barriers}
\end{figure}

\subsection{Usability Barriers}

The usability barriers reported in the user survey are depicted in
Fig.~\ref{fig:barriers} and aggregated into the following three categories:

\begin{enumerate}[leftmargin=1.4em,itemsep=4pt,topsep=4pt]
\item \textbf{Host--sandbox file sharing (UB-1):} How cumbersome is it to share files between the host and the sandbox? This includes regular files, dotfiles and the clipboard.
\item \textbf{Networking (e.g., ports, localhost, internet access) (UB-2):} How hard is it to get ports, localhost, and networking to work correctly?
\item \textbf{Environment provisioning (UB-3):} Does the environment (e.g., dotfiles, shell, tools, logins) have to be recreated every time in the sandbox? This includes ease of first use, custom Claude skill transfer, persisting configurations and package installations across restarts---as well as the initial setup of the sandboxes.
\end{enumerate}

Beyond these three usability barriers, users also noted usability issues regarding installation, multisandbox management, reliability and performance. However, to keep implementation realistic within the time constraints of the practicum, I only based my development and evaluation on the top three most common usability barrier categories. For the evaluation, each usability barrier category is broken down into capability tests in Section~\ref{sec:evaluation}.

\section{SideKernel}
\label{sec:sidekernel}

\subsection{Architecture}
\label{sec:architecture}

\begin{figure}[!t]
    \centering
    \includegraphics[width=1\columnwidth]{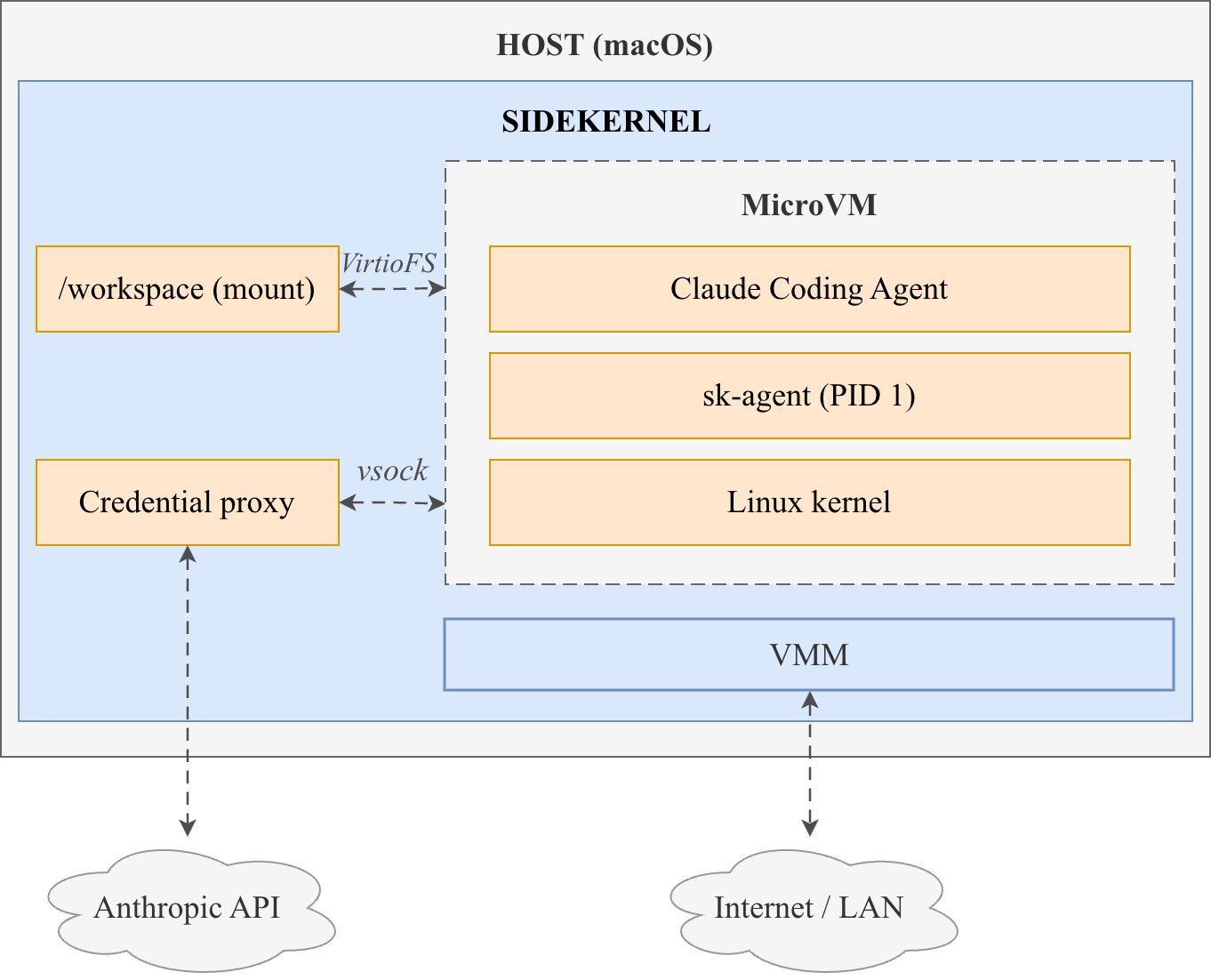}
    \caption{SideKernel's Architecture}
    \label{fig:architecture}
\end{figure}

A depiction of SideKernel's architecture can be found in Fig.~\ref{fig:architecture}. SideKernel consists of a virtual machine monitor (VMM) written in Swift on top of Apple's Virtualization framework. It uses a stripped-down precompiled Linux kernel by Kata Containers~\cite{kata}, over a curated Ubuntu root filesystem built on first run, both pinned down to their \texttt{SHA256} hashes.

Moreover, a core component of SideKernel is a minimal zero-dependency Rust agent that runs as PID 1 inside the microVM. This Rust agent is responsible for mounting the host's filesystem using VirtioFS, and for automatic port detection \& forwarding. The Rust agent uses virtual sockets (vsock) for communication with the host and exposes a pseudoterminal.

An important design decision is a small credential proxy server on the host, outside the sandbox. It intercepts the agent's API requests and injects AI coding agent credentials, so that no credentials are ever available inside the sandbox\footnote{The only exception is the first authentication inside the sandbox, where SideKernel reads the credentials and immediately deletes them from the sandbox while storing them in the macOS Keychain.}

\subsection{Design Goals and Implementation}

The user survey signals that usability issues constitute barriers to adoption of sandboxes for AI coding agents. Therefore, SideKernel is opinionated, and its design is guided by the usability issues the survey identified. This opinionated design means that the standard deny-all-by-default principle is applied selectively rather than absolutely, in favor of usability.
 
\begin{table*}[!tb]
\centering
\caption{SideKernel design goals traced to the usability barriers of Section~\ref{sec:survey} and the boundaries of Section~\ref{sec:threatmodel}}
\label{tab:designgoals}
\setlength{\tabcolsep}{3pt}
\renewcommand{\arraystretch}{1.15}
\footnotesize
\rowcolors{2}{rowalt}{}
\begin{tabularx}{\textwidth}{@{}l l >{\raggedright\arraybackslash}X >{\raggedright\arraybackslash}X c >{\raggedright\arraybackslash}X@{}}
\toprule
\textbf{UB} & \textbf{Boundary} & \textbf{Design Goal} & \textbf{Mechanism} & \textbf{Implemented} & \textbf{Risk Description} \\
\midrule
UB-1 & South & The current folder is the sandbox & Implicit bidirectional current directory mount with \texttt{VirtioFS} & \yes & The agent may read, edit or destroy anything in the mounted directory \\
UB-1 & South & External (non-mounted) files or directories reachable at runtime, with explicit consent & Command \texttt{sk-drop} with host-side approval. Drag and drop works into SideKernel Claude's sessions & \yes & --- \\
UB-1 & South & Copy and paste works between the host and SideKernel's Claude sessions & Bidirectional text (and image) clipboard & \yes & \NA \\
UB-2 & South & Exposed sandbox ports are \textit{automatically} available on the host & SideKernel auto-detects and forwards guest ports to \texttt{127.0.0.1} & \yes & A malicious agent can open ports to the host, exposing malicious services.  \\
UB-2 & North & Block all internet/external traffic, while the Claude agent still works & NAT kill-switch, Claude inference endpoints allowlisted by default & \yes & \NA \\
UB-3 & South & SideKernel feels like a second PC (e.g., installed packages and configurations persist across sandboxes) & Runtime OverlayFS persistence with a personalized SideKernel image, using the in-guest \texttt{save} command & \yes & \NA \\
UB-3 & South & One login authenticating on the Mac \textit{or} inside a sandbox authenticates both & The default Claude Code macOS Keychain entry is shared between the VMM of SideKernel (not the sandbox) and the host & \yes & \NA \\
UB-3 & South & Claude credentials are never exposed inside the sandbox & Credential proxy outside of the sandbox & \yes & \NA \\
UB-3 & South & Agent feels and behaves as configured on the host & Automatic skill, agent and plugin transfer & \yes & \NA \\
UB-3 & South & Arbitrary user secrets can be stored outside the sandbox, and injected securely & To be defined (Future Work, Section~\ref{sec:future}) & \no & \NA \\
UB-2 & North & The user can allowlist arbitrary destinations while egress is otherwise blocked & To be defined  (Future Work, Section~\ref{sec:future}) & \no & \NA \\
\NA & \NA & Agentic AI filesystem and process events inside the sandbox are logged &  To be defined (Future Work, Section~\ref{sec:future}) & \no & \NA \\
\NA  & North & External endpoints the agent contacted are logged & To be defined (Future Work, Section~\ref{sec:future}) & \no & \NA \\
\bottomrule
\end{tabularx}
\end{table*}

Each design goal of SideKernel relates to a usability barrier of Section~\ref{sec:survey}. The  capabilities developed bind to a specific usability barrier but some entail a deliberate risk acceptance, as described in SideKernel's threat model (Section~\ref{sec:threatmodel}) and summarized in Table~\ref{tab:designgoals}.

Table~\ref{tab:designgoals} also includes design goals not implemented for this paper. These goals around credentials management, observability of agentic events inside the sandbox, visibility of outbound connections, and more granular controls on the north boundary were scoped down from the initial aspirations of the practicum. Some of those features are present in other mature products such as Docker Sandboxes \cite{docker-sbx}.

The rest of this subsection describes the operation of SideKernel in more detail.

\subsubsection{Entering a sandbox} After installing SideKernel\footnote{Currently through a Makefile---in future work through brew or .dmg files}, a user can enter the shell of SideKernel by typing \texttt{sidekernel} (or the abbreviated alias \texttt{sk}). The command launches a new ephemeral microVM in the background and the current directory of the host is automatically mounted to this microVM.  While this implicit mount looks contrary to standard security practices, it is exactly what a Claude session (and other AI coding agents) does on the host, so it matches what the user expects. A user then can run \texttt{claude} inside SideKernel to launch Claude Code, as the latest version of Claude is preinstalled on the first launch. However, SideKernel also provides a faster way to start Claude. That is, by running \texttt{sclaude} directly on the host (which stands for \textit{Secure Claude} or \textit{SideKernel Claude}). 

\subsubsection{Authentication} If Claude is already authenticated on the host, running it inside SideKernel automatically provides an authenticated session. When running Claude inside SideKernel for the first time, SideKernel's VMM reads Claude's existing credentials from the host's macOS Keychain. This is only possible after an explicit approval from the user. The macOS Keychain access control prompt requests the user for their password and permission to allow SideKernel to write to it. If Claude is not authenticated on the host, authenticating Claude in the sandbox also auto-authenticates Claude to the host. In a nutshell, SideKernel's and the host's Claude share the same authentication.

\subsubsection{Sharing External Files} Host files and folders outside of the mounted directory can be requested from \emph{inside} SideKernel at runtime via an in-guest command, \texttt{sk-drop <host-path>}, which, combined with explicit user approval outside the sandbox, drops external files into SideKernel. SideKernel includes preloaded instructions for Claude on how to use \texttt{sk-drop}. Therefore, asking Claude for a macOS file, or dragging and dropping a file into a Claude session is enough to trigger \texttt{sk-drop} and make the file available to the agent. Moreover, the clipboard of the host is pastable into SideKernel. For SideKernel's Claude sessions particularly, images (e.g. clipboard screenshots) are pastable into the SideKernel session as well.

\subsubsection{Ports and networking} In SideKernel port forwarding is fully automatic. For example, running a Python server on port 4000 inside the sandbox makes that port accessible on the host without any extra steps or explicit port forwards. The Rust agent detects newly opened ports, and SideKernel auto-exposes them on the host's loopback address.

\subsubsection{Network (NAT) kill switch} SideKernel provides a network kill switch for the northbound channel, controlled by an in-guest command via \texttt{sk-net <on/off>}. Turning the kill switch on blocks all outbound traffic on the northbound channel. The only exception is the official API endpoint of Claude (\texttt{api.anthropic.com}), so that Claude can still run inference. Turning the kill switch off restores traffic on the northbound channel (this action requires explicit user approval). This capability is coordinated by SideKernel's Rust agent in communication with SideKernel's VMM. 

\subsubsection{Persistence and host parity} A user can use SideKernel as they would use any machine, install packages or change configurations, and then type the in-guest command \texttt{save} inside the sandbox. With save, a personalized image (an overlay filesystem) persists those changes across restarts and across sandboxes. Every time SideKernel is started or restarted, the VMM spins up and down a new ephemeral microVM in the background. Persisting this state gives the user a sense of continuity. To facilitate environment set-up and provisioning further, Claude's skills, plugins and agents are automatically copied in from the host before boot, so Claude in SideKernel works as transparently as possible.

\section{Evaluation: A Comparative Analysis}
\label{sec:evaluation}

The evaluation of SideKernel takes the form of a comparative analysis. First, I filter the list of discovered sandboxes against the inclusion requirements two to five (IR-2 to IR-5) as depicted in Fig.~\ref{fig:IR_filtering}. IR-1 (Open-Source) is deliberately excluded from the filtering. Restricting the study to open-source systems would exclude Docker Sandboxes~\cite{docker-sbx}, which is freely available and maintained by a well-known and reputable company in the systems industry. 

The results of the first stage of the filtering process are shown in detail in Table~\ref{tab:inclusion}. The sandboxes that pass the inclusion criteria and filtering process are Docker Sandboxes, Microsandbox, SmolVM, Shuru, VibeBox\footnote{VibeBox was included in the initial list for completeness. However it is an unmaintained project (the first public commit to VibeBox was on Feb 7, 2026 and last commit on Feb 18, 2026) which has only one contributor, its author. Moreover, after significant effort to install VibeBox on two separate macOS machines (macOS Tahoe 26.2, M1 \& M4 chips) I was unable to get a working sandbox, despite following its README.md instructions. Therefore, although the sandbox passes the inclusion criteria, it is currently not a suitable candidate for evaluation.} and, of course, SideKernel.

\subsection{Comparative Analysis Process}
\label{sec:comparative_analysis}

I break down each usability barrier into \textsc{capability tests}: A capability test is a \textit{specific sequence of unit actions}. Examples include (but are not limited to) running a host-side command, opening a new terminal window, dragging a file onto Claude's session window, pasting text or an image to the clipboard, entering a password etc. In Tables~\ref{tab:ub1}, \ref{tab:ub2}, and~\ref{tab:ub3}, \textbf{$T_n$} refers to the $n$-th capability test of each usability barrier category. The capability tests are summarized under each results table and are shown in detail in Appendix~\ref{app:tests}. The result of each capability test is then reported in two parts: 

\begin{enumerate}
    \item The \textit{binary outcome} of the capability test (mandatory)
    \item \textit{Friction notes} recording qualitatively the ease-of-use of \textit{passing} mechanisms. (optional, where applicable)
\end{enumerate}

In order to be as specific as possible in the qualitative descriptions I define and use the following terms:

\begin{enumerate}
    \item \textbf{Native:} The perceived user experience is exactly the same as on a macOS machine, requiring no additional unit actions
    \item \textbf{Frictionless:} The perceived user experience is pleasant, but not native-like, requiring one or two additional unit actions
    \item \textbf{(With) friction:} The perceived user experience is unpleasant, requiring three or more additional unit actions
\end{enumerate}

\subsection{UB-1: Host--Sandbox File Sharing}

Table~\ref{tab:ub1} summarizes the results of the capability tests for UB-1. Out of the nine (9) total tests, SideKernel passes eight (8), Docker Sandboxes passes six (6), and Microsandbox, Shuru, and SmolVM pass five (5). 

\ubtable{tab:ub1}{UB-1 --- Host--sandbox file sharing}{9}{\ubheaderA}{\ubbodyA}{1.5cm}{%
T1 & Mount read/create/delete, both directions \\
T2 & Mount mode propagation, both directions \\
T3 & Mount inode notification, host $\to$ sandbox \\
T4 & Unmounted file transfer, host $\to$ sandbox \\
T5 & Clipboard image, host $\to$ Claude in sandbox \\
T6 & Clipboard text, Claude in sandbox $\to$ host \\
T7 & Mount ownership propagation \\
T8 & Symlink resolution \\
T9 & Large file (10\,GB) mount integrity \\}

\smallskip\noindent\textbf{Friction Notes}\par

Docker Sandboxes and SideKernel offer a \textit{native} mounting experience automounting the CWD. The rest of the sandboxes require pre-launch mounts from the host causing \textit{friction}. In Microsandbox and SmolVM, permission bits (r, w, x) do not carry across
the south boundary. 

External file dropping (T4) is possible \textit{with friction} on Docker Sandboxes, Microsandbox, and SmolVM via a host-side command: the user must open a new terminal, recall the sandbox name, discover the sandbox path for mounting the file, and type/paste it. Each of these unit actions adds friction. Indicatively, the host-side commands are:

\begin{lstlisting}[style=cmds]
# Mounting
shuru run --mount <HOST-PATH>:<SBX-PATH>
msb run ubuntu --mount-dir <HOST-PATH>:<SBX-PATH>
smolvm machine start --name <n> -v <HOST-PATH>:<SBX-PATH>

# Copying external files
sbx cp <FILE> <SBX-NAME>:<SBX-PATH>
msb copy <FILE> <SBX-NAME> 
smolvm machine cp <FILE> <SBX-NAME>:<SBX-PATH> 
\end{lstlisting}

SideKernel's in-guest command \texttt{sk-drop <FILE>}, as well as the ability to drag and drop a file directly into the Claude session (and Claude executes \texttt{sk-drop} in the background) is \textit{frictionless}: requiring one in-guest command and approval on a macOS modal per external file.

A host Claude Code session allows pasting of both text and images in interactive mode~\cite{anthropic_claude_code_interactive_mode}. SideKernel is the only sandbox to pass the clipboard tests (T5, T6), supporting \textit{native-like} bidirectional text copy and paste, and unidirectional image pasting from the host clipboard into the SideKernel sandbox.

\newpage

\subsection{UB-2: Networking, Ports \& Localhost}

Table~\ref{tab:ub2} summarizes the results of the capability tests for UB-2. Out
of the five (5) total tests, Docker Sandboxes passes five (5), while Microsandbox, Shuru, SmolVM, and SideKernel pass three (3).

\ubtable{tab:ub2}{UB-2 --- Networking, ports \& localhost}{5}{\ubheaderB}{\ubbodyB}{1.5cm}{%
T1 & Runtime port forwarding \\
T2 & Restart port forwarding \\
T3 & DNS connection URL logs visibility \\
T4 & Egress control, blocking \\
T5 & Egress control, allowlisting \\}

\smallskip\noindent\textbf{Friction Notes}\par

All candidate sandboxes provide some way of port-forwarding, however they differ on the mechanism. SideKernel provides \textit{native-like} automatic runtime port-forwarding
without any user interaction; Docker Sandboxes allows \textit{frictionless}
manual runtime port forwarding through a host command.

Microsandbox, SmolVM and Shuru allow port-forwarding but not on a running sandbox \textsc{(T1)}. They require the ports to be defined on the host before starting the sandbox, which introduces \textit{friction} (especially if the developer does not know in advance if and which ports they might need open and exposed to the host). Indicatively, the commands are:

\begin{lstlisting}[style=cmds]
msb run ubuntu --port <HOST-PORT>:<SBX-PORT> 
smolvm machine <run|create> --port <HOST>:<SBX-PORT>
shuru run -p <HOST-PORT>:<SBX-PORT> -- nginx 
\end{lstlisting}

Docker Sandboxes is the only sandbox providing DNS observability on the north boundary (\texttt{T3}). All sandboxes but SideKernel provide granular north boundary security controls blocking egress traffic with allowlists. SideKernel provides an in-guest kill-switch without however the granularity other sandboxes provide.

Notably, Docker Sandboxes' allowlist is configurable through the sandbox's
built-in TUI beyond declarative host policies, making it easy to disallow or allow
connections during runtime. It continuously monitors DNS connections, lists all the
domains that the sandbox has communicated with, and allows the allowlist to be
toggled seamlessly through the terminal interface (or new records to be added).

These observability and firewall features are additional capabilities that are not normally present in the host; therefore, describing friction for those is not applicable.

\subsection{UB-3: Environment Provisioning \& Setup}
Table~\ref{tab:ub3} summarizes the results of the capability tests for UB-3. Out
of the nine (9) total tests, Docker Sandboxes and SideKernel pass eight (8), and SmolVM, Microsandbox and Shuru pass four (4).
\ubtable{tab:ub3}{UB-3 --- Environment provisioning \& setup}{9}{\ubheaderC}{\ubbodyC}{1.5cm}{%
T1 & Agent preinstallation \\
T2 & Credential inheritance \\
T3 & Agent state persistence \\
T4 & Configuration file persistence \\
T5 & Package installation persistence \\
T6 & Git state visibility \\
T7 & Live checkpoint persistence \\
T8 & Secret store, agent credentials \\
T9 & Secret store, custom secrets \\}

\smallskip\noindent\textbf{Friction notes.}\par

Docker Sandboxes and SideKernel are the only two sandboxes that come with preinstalled agents \textsc{(T1)}. However, Docker Sandboxes supports many major agents including Claude Code, Codex, Gemini CLI, GitHub Copilot, opencode and more. SideKernel currently only supports Claude Code. SideKernel is the only sandbox that shares agent authentication with the host, providing a \textit{frictionless} experience on the first run (entering the
macOS password once on the Keychain's access control prompt).

Configuration and package installation persistence is feasible in all sandboxes, although in Shuru, Microsandbox, and SmolVM \textit{with friction}. Docker Sandboxes and SideKernel however persist changes that survive restarts at runtime in a \textit{frictionless} manner.

With regard to AI agent configuration, SideKernel is the only sandbox that inherits Claude Code's skills, plugins and agents from the host providing a \textit{native-like} experience.
With regard to credentials management, Docker Sandboxes is the only sandbox with a built-in secret store for custom credentials.

\section{Discussion}
\label{sec:discussion}

\subsection{Interpreting the Aggregate Results}

Across all 23 capability tests, Docker Sandboxes and SideKernel pass 19 tests, SmolVM, Microsandbox and Shuru pass 12 tests. The combination of the aggregated test results with the qualitative friction notes indicates that Docker Sandboxes and SideKernel are the most usable sandboxes in the study (following the benchmark introduced in Section~\ref{sec:benchmark_construction}). However, this is only true inside the confines of the usability barriers that were tested and it is not a generalizable statement. It is also worth noting that even among AI coding agent sandboxes, the intended use varies. Docker Sandboxes, for example, is built for enterprise deployments and multi-sandbox orchestration. Shuru is intended to be used as a Claude skill invoked by an AI agent running on the host, self-isolating its tool runs inside the sandbox. SideKernel attempts to replicate as closely as possible the experience a user has on the host, from inside a microVM. Last but not least, Microsandbox and SmolVM are hybrid sandboxes, focusing primarily on SDK programmable agents and providing CLI features as a sidecar rather than as their main design.

\subsection{Comparison with Docker Sandboxes}

SideKernel provides a few features bound to usability barriers of the user study that no other sandbox does: \textit{frictionless} external file drops through an in-guest command or drag and drop into Claude, automatic port forwarding of newly opened ports inside the sandbox, bidirectional clipboard capabilities inside Claude sessions (e.g., pasting images), and synchronized authentication with the host. These are small but not insignificant contributions to the current market of sandboxes. Moreover, I intend to open-source SideKernel, a fact differentiating it from the free but closed-source Docker Sandboxes.

That said, Docker Sandboxes is an undeniably more feature-rich sandbox, including advanced observability and auditing capabilities, governance and granular policies, a versatile terminal user interface, easy multi-sandbox management features, mature credential management, and others that were out of scope for this study. However, within the confines of the usability barriers that the user study discovered, I conclude Docker Sandboxes and SideKernel are the most capable sandboxes in regard to usability features---something which a reader can evaluate through the capability tests and the friction notes. 

\subsection{Real-World Considerations}

\begin{itemize}
    \item SideKernel is designed for single macOS machines, but tested only on M1 and M4 chips and only on the latest Tahoe 26.2 OS version. It might still include undetected bugs and rough edges. 
    \item The security of SideKernel is implied from an architectural perspective, but the implementation has not undergone a formal security review.
    \item SideKernel is not yet notarized (and it is self-signed). It currently has no observability out of the box.
    \item SideKernel currently provides no enterprise features, including logging, declarative policies and centralized governance capabilities. 
\end{itemize}

\newpage
\section{Limitations and Threats to Validity}
\label{sec:limitations}

Below I describe all the limitations and threats to the validity of the evaluation and development of SideKernel. The majority of those limitations are attributed to the time constraints of the practicum (which required scoping down), and single-evaluator bias.

\begin{itemize}[leftmargin=1.4em,itemsep=4pt,topsep=4pt]
\item \textbf{Evaluator bias.} The comparative analysis unavoidably introduces bias, because the developer is also the evaluator. Although I took every reasonable measure to reduce it (e.g. fixed success criteria, binary outcomes based on built-in capabilities,
auditable tests, standardized and transparent friction notes), a better
methodology would involve multiple users, something, however, practically
infeasible within the practicum's timeline.
\item \textbf{User survey instrument bias.} The user survey instrument is designed by the developer. Many of the survey's questions are multiple-choice or multi-select format. This anchors respondents towards the pre-formulated options, which introduces bias. I mitigate this by providing a free-text ``Other'' option on closed questions, as well as enriching the instrument with open free-text questions. While this likely reduces the instrument bias, it does not eliminate it.
\item \textbf{Sampling bias.} The online survey's sample tilts towards a security-aware audience. Real-world AI coding agent sandbox adoption is therefore probably significantly lower than the 38.2\% that the study records.
\item \textbf{Sample size.} While the sample (N\,=\,80 respondents; 33 current or former sandbox users) is likely sufficient for formative/directional signals, it is low. The results were used for informed development decisions and a framework for the comparative analysis, rather than statistically generalizable data.
\item \textbf{Informal study status.} The formative user survey was conducted as informal practicum research and did not undergo a formal Georgia Tech IRB review. The study is not designed to produce generalizable knowledge but rather its results are interpreted as signals for the development and evaluation of SideKernel, rather than rigorous academic data.
\item \textbf{Usability is measured indirectly.} The tests measure how well a sandbox addresses the survey's usability barriers, not usability in any general sense. 
\item \textbf{Binary scoring without gradation.} The binary score of the capability tests is blind to friction. While I remediate this with qualitative notes on friction, it does not eliminate the fact that the score itself does not weigh the very nuances that describe usability.
\item \textbf{Binary scoring without weights.} The binary score of the capability tests does not consider which capabilities are important for the users in practice. The scores are naive, in the sense that the binary outcomes are unweighted. Therefore, the results are approximations. 
\item \textbf{Platform constraints.} SideKernel is a Linux microVM on top of macOS. It therefore cannot offer capabilities that Seatbelt-based sandboxes can. For example, it is unsuitable for testing native macOS SwiftUI applications, and certain macOS terminal commands are unavailable inside SideKernel. 
\item \textbf{Scoping constraints.} SideKernel is a microVM sandbox. Sandboxes based on other isolation primitives such as macOS Seatbelt might also provide sufficient security guarantees---depending on the use case and threat model. SideKernel is not evaluated against such sandboxes which were deliberately left out of scope to keep the project manageable within the time constraints of the practicum. 
\end{itemize}

\section{Future Work}
\label{sec:future}

The comparative analysis exposed areas for improvement in SideKernel, in
particular capabilities that similar sandboxes already implement. Future work
will include replacing the NAT kill-switch with a full network proxy server that
provides granular allowlisting. Such a network proxy will also allow implementation of DNS observability. On the south boundary, the remaining work is a secure generic credential management store, beyond Claude's credentials. Finally, SideKernel will support more major agents (in a similar way to how Claude Code is currently integrated), including Codex, Microsoft Copilot
and others.

\section{Conclusion}
\label{sec:conclusion}

SideKernel constitutes a capable local microVM sandbox for AI coding agents that addresses many of the usability barriers reported in the user study. It is currently the only local microVM sandbox that solves the bidirectional clipboard issue in Claude Code, provides automatic runtime port forwarding, integrates Claude Code's native macOS Keychain authentication seamlessly, and transfers the host's Claude skills and agents into the sandbox automatically. Moreover, it provides a frictionless way to drop external files into the sandbox, and eases environment provisioning through a personalized image.

Another highly capable microVM sandbox in the context of the usability barriers
addressed is Docker Sandboxes. It is the only sandbox that allows or blocks DNS
requests through a TUI, provides custom credential management, preinstalls most major AI coding agents, and offers observability on top.

Microsandbox, Shuru and SmolVM follow, with more limited AI coding agent capabilities.

I attribute this to the fact that there has been little serious effort in the industry focused on AI coding agents for local macOS environments. The arms race pushes sandboxes towards autonomous programmable SDK agents, leaving coding agents as an afterthought, despite this being a demonstrably real security problem. This part of the industry is still young and rapidly developing. The fact that an infrastructure company as big as Docker did not release sbx until January 30, 2026~\cite{docker-sbx} further strengthens this claim, and justifies the development of SideKernel.

A personal note: I already use SideKernel daily. I now develop SideKernel inside of SideKernel!

\clearpage
\onecolumn
\linespread{1.15}\selectfont

\makeatletter
\def\section{\@startsection{section}{1}{\z@}{1.5ex plus 1.5ex minus 0.5ex}%
{0.7ex plus 1ex minus 0ex}{\normalfont\normalsize\scshape}}%
\makeatother

\appendices

\section{Full Task Definitions}
\label{app:tests}

This appendix gives the complete definition of every task referenced in Section~\ref{sec:evaluation}. Each task is a capability test with a binary outcome, scored according to the criteria in Section~\ref{sec:benchmark_construction}.

\begingroup
\footnotesize
\renewcommand{\arraystretch}{1.25}
\setlength{\tabcolsep}{4pt}

\rowcolors{2}{rowalt}{}
\begin{longtable}{@{}p{0.8cm} >{\raggedright\hyphenpenalty=10000\exhyphenpenalty=10000\arraybackslash}p{42mm} >{\raggedright\arraybackslash}p{112mm}@{}}
\caption{UB-1 --- Host--sandbox file sharing: full task definitions}\label{tab:defub1}\\
\toprule
\textbf{ID} & \textbf{Name} & \textbf{Definition} \\
\midrule
\endfirsthead
\multicolumn{3}{@{}l}{\footnotesize\itshape Table~\ref{tab:defub1} (continued)}\\
\toprule
\textbf{ID} & \textbf{Name} & \textbf{Definition} \\
\midrule
\endhead
\bottomrule
\multicolumn{3}{@{}p{155mm}@{}}{\footnotesize\raggedright
$^{a}$~The goal is to let the user read any external file easily and securely. Easily means from inside the sandbox through a built-in mechanism, without sandbox restarts or host-side commands; securely means with explicit user approval.}\\
\endlastfoot
\tdef{T1}{Mount CRUD, both directions}{Read, create and delete the mount dir file \texttt{mount/workspace-file.txt} and validate bidirectionally all actions.}
\tdef{T2}{Mount mode propagation, both directions}{Run \texttt{chmod 777 mount/shell.sh}. Test that all the permission bits (r, w, x) carry across.}
\tdef{T3}{Mount inode notification, host $\to$ sandbox}{Run \texttt{inotifywait} on \texttt{workspace-file.txt} inside the sandbox, then edit the file on the host. Test that the notification fired.}
\tdef{T4}{Unmounted file transfer, host $\to$ sandbox}{Test whether \texttt{<HOST-PATH>/external-file.txt} can be accessed and read inside the sandbox, without host-side commands, without sandbox restart, and with explicit approval.$^{a}$}
\tdef{T5}{Clipboard image, host $\to$ Claude in sandbox}{Copy an image to the host clipboard. Test that it pastes into the Claude session inside the sandbox.}
\tdef{T6}{Clipboard text, sandbox $\to$ host}{Copy text from the Claude session inside the sandbox. Test that it pastes on the host.}
\tdef{T7}{Mount ownership propagation}{Create \texttt{mount/sandbox-created.txt} inside the sandbox. Test that \texttt{ls -l} on the host shows the host user as owner.}
\tdef{T8}{Symlink resolution}{Create a symlink in \texttt{mount/} to \texttt{workspace-file.txt}. Test that it resolves.}
\tdef{T9}{Large file mount integrity}{Mount the zero-filled \texttt{10GB <HOST-PATH>/big-file.txt} to the sandbox. Check that it mounts successfully and that the \texttt{md5sum} matches the one on the host.}
\end{longtable}

\rowcolors{2}{rowalt}{}
\begin{longtable}{@{}p{0.8cm} >{\raggedright\hyphenpenalty=10000\exhyphenpenalty=10000\arraybackslash}p{42mm} >{\raggedright\arraybackslash}p{112mm}@{}}
\caption{UB-2 --- Networking, ports \& localhost: full task definitions}\label{tab:defub2}\\
\toprule
\textbf{ID} & \textbf{Name} & \textbf{Definition} \\
\midrule
\endfirsthead
\multicolumn{3}{@{}l}{\footnotesize\itshape Table~\ref{tab:defub2} (continued)}\\
\toprule
\textbf{ID} & \textbf{Name} & \textbf{Definition} \\
\midrule
\endhead
\bottomrule
\endlastfoot
\tdef{T1}{Runtime port forwarding}{Run \texttt{python3 -m "http.server" 5000} inside the sandbox. Configure port forwarding without restarting the sandbox for 5000. Test that \texttt{curl localhost:5000} is reachable from the host.}
\tdef{T2}{Restart port forwarding}{Run \texttt{python3 -m "http.server" 6000} inside the sandbox. Test that \texttt{curl localhost:6000} is reachable from the host after explicitly configuring port forwarding.}
\tdef{T3}{DNS connection URL logs visible}{Provides a DNS network log panel listing the websites connected to.}
\tdef{T4}{Egress control, blocking}{Test that blocking outbound internet access is feasible with a built-in sandbox mechanism enforced by the hypervisor. Run the sandbox and block all traffic.}
\tdef{T5}{Egress control, allowlisting}{If blocking outbound internet access is feasible, test whether allowlisting is feasible natively, enforced by the hypervisor. Run the sandbox and allowlist \texttt{curl https://example.com}.}
\end{longtable}

\rowcolors{2}{rowalt}{}
\begin{longtable}{@{}p{0.8cm} >{\raggedright\hyphenpenalty=10000\exhyphenpenalty=10000\arraybackslash}p{42mm} >{\raggedright\arraybackslash}p{112mm}@{}}
\caption{UB-3 --- Environment provisioning \& setup: full task definitions. For each task, the tester checkpoints the sandbox (if applicable)}\label{tab:defub3}\\
\toprule
\textbf{ID} & \textbf{Name} & \textbf{Definition} \\
\midrule
\endfirsthead
\multicolumn{3}{@{}l}{\footnotesize\itshape Table~\ref{tab:defub3} (continued)}\\
\toprule
\textbf{ID} & \textbf{Name} & \textbf{Definition} \\
\midrule
\endhead
\bottomrule
\endlastfoot
\tdef{T1}{Agent preinstallation}{Test that \texttt{claude --version} succeeds on first boot, without the tester installing it manually.}
\tdef{T2}{Credential inheritance}{Ensure Claude is authenticated on the host. Test that the agent completes a prompt--response round-trip inside the sandbox without the tester performing a login in the sandbox.}
\tdef{T3}{Agent state persistence}{Restart the sandbox. Test that Claude is still present and authenticated.}
\tdef{T4}{Configuration file persistence}{Define alias \texttt{print\_hi} in the sandbox shell rc, and restart it. Test that \texttt{print\_hi} resolves.}
\tdef{T5}{Package installation persistence}{Run \texttt{pip install -r requirements.txt}, and restart it. Test that \texttt{python -c "import numpy"} exits successfully.}
\tdef{T6}{Git state visibility}{Run \texttt{git init .} on the host and create a new file called \texttt{git.txt}. Enter the sandbox and run \texttt{git diff .}. Test whether \texttt{git.txt} appears on the diff.}
\tdef{T7}{Live checkpoint persistence}{Test whether the sandbox supports \textbf{live checkpoints}. A live checkpoint saves the \textit{running disk state} of a sandbox, including installed packages and configuration changes. Run \texttt{apt-get install nmap}, checkpoint and restart the sandbox, then verify whether \texttt{nmap} is still installed.}
\tdef{T8}{Secret store, agent credentials}{Test whether the sandbox provides a designated secret store (e.g., macOS Keychain) for Claude credentials (OAuth/API Keys).}
\tdef{T9}{Secret store, custom secrets}{Test whether the sandbox provides a designated secret store (e.g., macOS Keychain) for arbitrary custom secrets.}
\end{longtable}
\endgroup

\section{Google Survey Results}
\label{app:survey}

Table~\ref{tab:survey} reports the complete survey results. Questions are presented for the relevant corresponding groups, for example, sandboxing tools previously used are over the group of former users. Reasons for never using a sandbox are over the group of non-users etc. For brevity, I omit free-text responses/options that were only selected by a single respondent.

\begingroup
\footnotesize
\setlength{\tabcolsep}{0pt}
\setlength{\LTleft}{\fill}\setlength{\LTright}{\fill}
\renewcommand{\arraystretch}{0.97}
\newcommand{\qq}[1]{\multicolumn{3}{@{}>{\columncolor{qband}}l@{}}%
  {\rule{0pt}{2.4ex}\hspace*{2pt}\textbf{#1}}\\[1.5pt]}
\newcommand{\rr}[1]{\hspace*{1.6em}#1}

\rowcolors{2}{rowalt}{}
\begin{longtable}{@{}>{\hspace*{2mm}}l >{\hspace*{10mm}}r >{\hspace*{10mm}}r<{\hspace*{2mm}}@{}}
\caption{Complete Google Survey results}\label{tab:survey}\\
\toprule
\textbf{Question / Response} & \textbf{n} & \textbf{\%} \\
\midrule
\endfirsthead
\multicolumn{3}{@{}l}{\footnotesize\itshape Table~\ref{tab:survey} (continued)}\\
\toprule
\textbf{Question / Response} & \textbf{n} & \textbf{\%} \\
\midrule
\endhead
\midrule
\multicolumn{3}{r@{}}{\footnotesize\itshape continued on next page}\\
\endfoot
\bottomrule
\multicolumn{3}{@{}l@{}}{\footnotesize\textsuperscript{$\dagger$}~Offered to current users only; not applicable to the combined percentage base.}\\
\multicolumn{3}{@{}l@{}}{\footnotesize\textsuperscript{$\ddagger$}~Offered to former users only; not applicable to the combined percentage base.}\\
\endlastfoot

\qq{Q1. Do you currently use AI coding agents? ($N=80$)}
\rr{Yes} & 68 & 85.0 \\
\rr{No} & 12 & 15.0 \\

\qq{Q2. Professional role ($n=68$)}
\rr{Cybersecurity (Engineering, OffSec etc)} & 23 & 33.8 \\
\rr{Software developer or engineer} & 14 & 20.6 \\
\rr{Platform engineer} & 10 & 14.7 \\
\rr{Technical manager or product owner} & 8 & 11.8 \\
\rr{Hobbyist developer} & 4 & 5.9 \\
\rr{Student} & 4 & 5.9 \\
\rr{Data/AI engineer or scientist} & 3 & 4.4 \\
\rr{Other (e.g., founder)} & 2 & 2.9 \\

\qq{Q3. AI coding agents used (multi-select, $n=68$)}
\rr{Claude Code} & 59 & 86.8 \\
\rr{OpenAI Codex} & 26 & 38.2 \\
\rr{GitHub Copilot (agent mode/chat, not just inline suggestions)} & 23 & 33.8 \\
\rr{Cursor} & 13 & 19.1 \\
\rr{Gemini CLI} & 10 & 14.7 \\
\rr{opencode} & 2 & 2.9 \\

\qq{Q4. Operating systems used (multi-select, $n=68$)}
\rr{macOS} & 41 & 60.3 \\
\rr{Linux} & 37 & 54.4 \\
\rr{Windows} & 22 & 32.4 \\
\rr{Windows (in WSL)} & 13 & 19.1 \\

\qq{Q5. Have you used an isolation environment for AI coding agents? ($n=68$)}
\rr{Yes, I currently use one} & 26 & 38.2 \\
\rr{Yes, but I stopped using it} & 7 & 10.3 \\
\rr{No, I have never used one} & 35 & 51.5 \\

\qq{Q6. Sandboxing tools in use (multi-select, current users, $n=26$)}
\rr{Docker Sandboxes} & 12 & 46.2 \\
\rr{Full VM (UTM, Parallels, Lima etc.)} & 9 & 34.6 \\
\rr{Nono.sh} & 7 & 26.9 \\
\rr{Separate machine dedicated for AI (e.g., Mac Mini)} & 2 & 7.7 \\

\qq{Q7/Q10. Barriers experienced (multi-select, current + former users, $n=33$)}
\rr{No significant difficulties\textsuperscript{$\dagger$}} & 9 & n/a \\
\rr{Had to set everything up from scratch (dotfiles, shell, tools, logins)} & 8 & 24.2 \\
\rr{Getting ports / localhost / networking to work is difficult} & 8 & 24.2 \\
\rr{Sharing files between sandbox and host is hard or slow} & 7 & 21.2 \\
\rr{Security benefit not worth the effort\textsuperscript{$\ddagger$}} & 6 & n/a \\
\rr{Managing multiple sandboxes is confusing} & 5 & 15.2 \\
\rr{Initial installation / setup is too hard} & 5 & 15.2 \\
\rr{It is slow, or slows down my machine} & 4 & 12.1 \\
\rr{The sandbox itself crashes or breaks} & 3 & 9.1 \\
\rr{OS mismatch (e.g., macOS host, Linux-only sandbox)} & 3 & 9.1 \\
\rr{Usual host tools do not work inside the sandbox} & 2 & 6.1 \\

\qq{Q9. Sandboxing tools previously used (multi-select, former users, $n=7$)}
\rr{Docker Sandbox} & 3 & 42.9 \\
\rr{Full VM (UTM, Parallels, Lima etc.)} & 2 & 28.6 \\
\rr{Microsandbox} & 1 & 14.3 \\
\rr{Separate machine dedicated for AI (e.g., Mac Mini)} & 1 & 14.3 \\
\rr{OrbStack} & 1 & 14.3 \\
\rr{OpenHands with Docker} & 1 & 14.3 \\

\qq{Q12. Reasons for never using a sandbox (multi-select, $n=35$)}
\rr{I have never considered it} & 21 & 60.0 \\
\rr{I am not aware of any good options} & 11 & 31.4 \\
\rr{The security benefit does not seem worth the effort for my use case} & 9 & 25.7 \\
\rr{The built-in sandboxing in my coding agent is sufficient} & 7 & 20.0 \\
\rr{I looked into it, but it seemed too difficult} & 4 & 11.4 \\
\end{longtable}
\endgroup

\twocolumn
\linespread{1.0}\selectfont
\makeatletter
\def\section{\@startsection{section}{1}{\z@}{3.0ex plus 1ex minus 0.3ex}%
{0.7ex plus 0.3ex minus 0.2ex}{\normalfont\normalsize\centering\scshape}}
\makeatother
\section*{Acknowledgment}
I co-developed the SideKernel codebase with AI (Claude, Anthropic).
I also used AI to refine the grammar and phrasing of this paper, to
generate the LaTeX code for tables and the draw.io code for diagrams,
and to review the manuscript for factual errors. AI was used in a
supporting capacity throughout every task, but the substance of the work is my own,
including but not limited to the text of this paper, design decisions, threat model,
methodology, instrument design, content of the tables and diagrams,
evaluation, analysis, research, and conclusions.
\bibliographystyle{IEEEtranN} 
\bibliography{references}

\end{document}